\documentclass[runningheads]{svmult}

\usepackage{makeidx}   
\usepackage{graphicx}  
\usepackage{subeqnar}  
\usepackage{multicol}  
\usepackage{physprbb}  
\makeindex             

\begin{document}
\title*{The Chemistry of Extragalactic Globular Clusters}
\titlerunning{The Chemistry of Extragalactic Globular Clusters}
\author{M.Kissler-Patig\inst{1}
\and T.H.Puzia \inst{2}
\and R.Bender\inst{2}
\and P.Goudfrooij\inst{3}
\and M.Hempel\inst{1}
\and C.Maraston\inst{2}
\and T.Richtler\inst{4}
\and R.Saglia\inst{2}
\and D.Thomas\inst{2}
}
\authorrunning{Kissler-Patig et al.}
%
%
\institute{European Southern Observatory, Germany
\and Sternwarte der LMU M\"unchen, Germany
\and Space Telescope Science Institute, USA
\and Universidad de Concepci\'{o}n, Chile}

\maketitle              

\begin{abstract}
We present preliminary results of VLT/FORS spectroscopy of globular
clusters in nearby early-type galaxies. Our project aims at studying the
chemistry and determine the ages of globular cluster (sub-)populations. 
First results indicate that
the different galaxies host from little to significant intermediate-age
populations, and that the latter have $\alpha$-element over iron ratios
closer to solar than the old population that show an $\alpha$-element 
enhancement similar to the diffuse stellar light.
\end{abstract}

\section{The Project}
Our group (P.I.: M.Kissler-Patig) started a campaign to obtain high
S/N spectra of representative globular clusters in early-type galaxies.
This work will form the bulk of Thomas H.~Puzia's PhD thesis and be used in
combination with optical/NIR photometry in Maren Hempel's PhD thesis.

So far, we have obtained intermediate-resolution spectra 
(R$\sim$800) of globular clusters in six early-type galaxies using
the FORS2/MXU instrument attached to ESO's VLT. 
These data are complemented by spectra 
of the host galaxy's integrated light out to large radii 
(2-3 R$_{eff}$). Our sample covers galaxies with various properties:
from bright ($M_B<-20.5$ mag) and faint ($M_B>-20.5$ mag) galaxies,
located in the field, in groups, and in clusters. 

The aim is to study the chemical composition of globular clusters
in order to understand the formation and evolution of their host
galaxies. In particular, we are investigating age differences within a
galaxy, are looking for differences in ages/metallicities between bright/faint 
and cluster/field galaxies; and are comparing the line indices of the
globular clusters with the ones for the corresponding integrated light
at comparable radii. 

Here, we presents first results for 4 galaxies:

\begin{center}
\begin{tabular}{lcccc}
\hline
Galaxy    & Type &  M$_B$    & $\rho$  & (m-M) \\
\hline
NGC 1380  & S0   & -20.04  & 1.54 & 31.23 \\
NGC 2434  & E0   & -19.48  & 0.19 & 31.67 \\
NGC 3115  & S0   & -19.18  & 0.09 & 29.93 \\
NGC 3379  & E1   & -19.39  & 0.52 & 30.12 \\
\hline
\end{tabular}
\end{center}


\section{The Balmer lines}

The best age indicators in the Lick system remain the Balmer lines.
Among these, H$\beta$ is mostly used, but is super-seeded in diagnostic
quality by H$\gamma_A$ for good data (see Puzia's contribution in these
proceedings).

The Balmer line needs to be plotted against an $\alpha$-element
insensitive index (e.g.~[MgFe]) otherwise any derived age is a function
of the $\alpha$-enrichment of that particular globular cluster.

\begin{figure}[ht]
\begin{center}
\includegraphics[width=.95\textwidth]{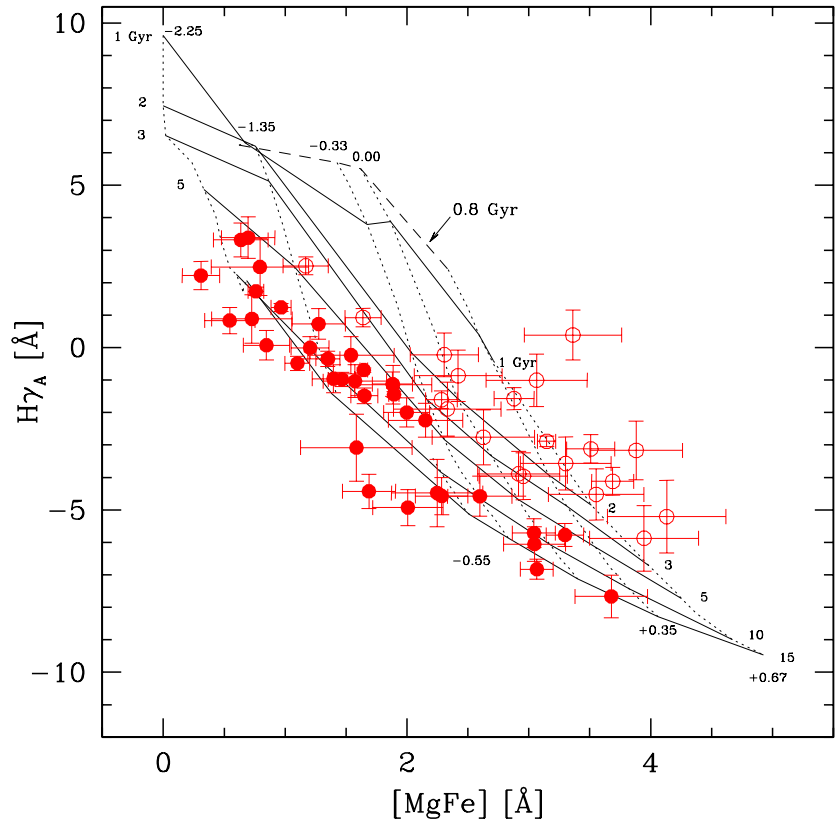}
\end{center}
\caption[]{H$\gamma_A$ plotted against [MgFe]}
\label{fig_age}
\end{figure}

Fig.~\ref{fig_age} shows H$\gamma_A$ plotted against [MgFe] for all studied 
globular clusters in the four galaxies. We compare our data to 
the latest SSP model grid of Maraston et al.~(2002) for ages 0.8--15 Gyr and 
metallicities $-2.25$ to +0.6 dex. Open symbols represent globular
clusters formally younger than 3 Gyr.


\section{The Metals}

Using the Lick indices, we can probe 1) the total metallicity, best
traced by [MgFe], a combination of Mg and two Fe indices that result in
a fairly $\alpha$-element independent quantity; and 2) the
$\alpha$-element ratio, well traced by comparing Fe and Mg.

The total metallicity traces the overall chemical enrichment
during the star formation history of the galaxy. As expected from
photometric studies, we observe in all galaxies a wide spread in
metallicity ranging from [Fe/H]$\sim -2.5$ dex to solar or above.

\begin{figure}[ht]
\begin{center}
\includegraphics[width=.75\textwidth]{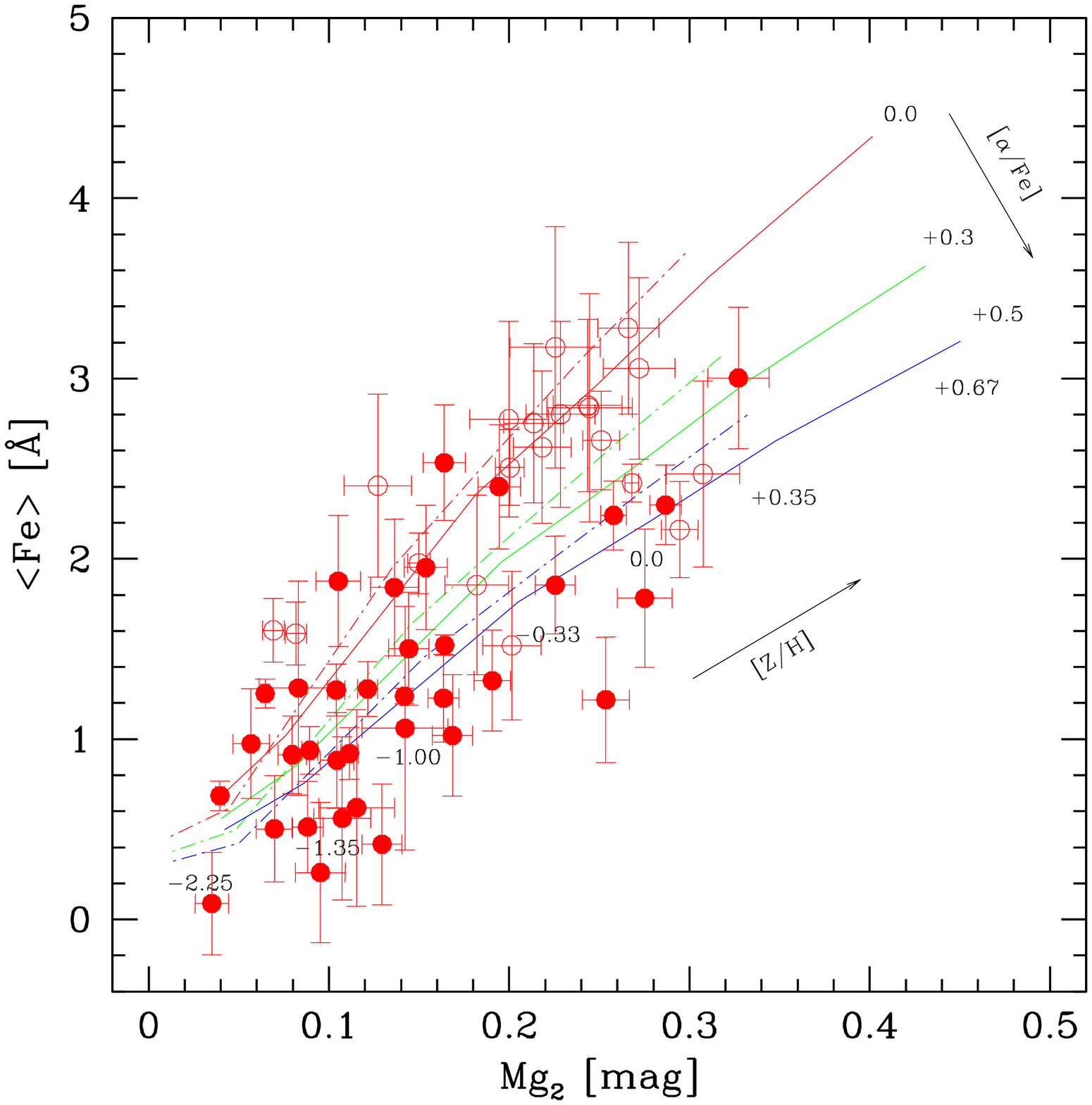}
\end{center}
\caption[]{Mg$_2$ plotted against $<$Fe$>$}
\label{fig_met}
\end{figure}

Fig.~\ref{fig_met} shows Mg$_2$ plotted against $<$Fe$>$ as diagnostic for
$\alpha$-element enrichment. The Mg$_2$ index traces the strength of the Mg
($\alpha$ element) absorption feature at 5180 \AA . The $<$Fe$>$ index is a 
composite index of two line indices Fe5335 and Fe5270 which trace the
strength of strong iron lines at around 5300 \AA . The iso-metallicity lines 
show SSP model predictions computed for three $\alpha$/Fe ratios (solar, 
$+0.3$ dex, $+0.5$ dex) assuming a 13 Gyr old stellar population (Thomas et 
al.~2002). The lines range from -2.25 dex (lower left) up to +0.67 dex (upper
right) in metallicity. Open symbols show clusters formally younger than
3 Gyr (see Fig.~\ref{fig_age}).


\section{The Results}

\underline{AGES:}

\begin{itemize}

\item The most striking feature of the age-metallicity plot is the
fact that the globular clusters with metallicities below [Fe/H]$\sim -0.8$ 
span a very narrow age range around the $\sim$12 Gyr isochrone, i.e.~comparable
to Milky Way halo clusters. For metallicities above [Fe/H]$\sim -0.8$
dex, the globular clusters show a large spread in ages over several Gyr.
In particular, a significant number of intermediate-age clusters appear
to be present (the ratio old to intermediate varying from galaxy to
galaxy). A caveat to the latter interpretation might be the
influence of fluctuations of the horizontal branch morphology. However,
the spectroscopic intermediate-ages seem confirmed by optical-NIR
photometry (Puzia et al.~2002, Hempel et al.~2002, see also Hempel et
al.~in these proceedings). 

\end{itemize}

\noindent\underline{METALLICITY:}

\begin{itemize}

\item The clusters in all galaxies span a similar metallicity
range as in the Milky Way. We found neither super metal-poor nor obvious super
metal-rich clusters.

\item In all galaxies, the metal-rich clusters ([Fe/H]$>-1.00$ dex) span a wide 
range of [$\alpha$/Fe] ratios indicating a large variety of formation time 
scales within a single galaxy. 
Interestingly, the intermediate-age globular clusters appear to have, on
average, lower [$\alpha$/Fe] values than the old clusters. While the bulk
of the intermediate-age clusters has ratios close to solar, the old
metal-rich clusters tend to be $\alpha$-element enriched, similar the
diffuse stellar light of the host galaxies.

\end{itemize}


\end{document}